\documentclass[runningheads]{llncs}

\usepackage[T1]{fontenc}
\PassOptionsToPackage{prologue,dvipsnames}{xcolor}
\usepackage{bbold}
\usepackage{amsmath, amssymb, stmaryrd}
\usepackage{varioref}
\usepackage{hyperref}
\usepackage{cleveref}
\usepackage{mathpartir}
\usepackage{todonotes}
\usepackage{xspace}
\usepackage{nccmath}

\usepackage{tikz}
\usetikzlibrary{cd}
\usepackage{ifmtarg}
\usepackage{float}

\usepackage{esvect}
\usepackage{tabulary}

\usepackage{contour}
\usepackage[normalem]{ulem}

\contourlength{0.8pt}

\usepackage{adjustbox}

\DeclareMathSymbol{\Gamma}{\mathalpha}{operators}{0}
\DeclareMathSymbol{\Delta}{\mathalpha}{operators}{1}
\DeclareMathSymbol{\Theta}{\mathalpha}{operators}{2}
\DeclareMathSymbol{\Lambda}{\mathalpha}{operators}{3}
\DeclareMathSymbol{\Xi}{\mathalpha}{operators}{4}
\DeclareMathSymbol{\Pi}{\mathalpha}{operators}{5}
\DeclareMathSymbol{\Sigma}{\mathalpha}{operators}{6}
\DeclareMathSymbol{\Upsilon}{\mathalpha}{operators}{7}
\DeclareMathSymbol{\Phi}{\mathalpha}{operators}{8}
\DeclareMathSymbol{\Psi}{\mathalpha}{operators}{9}
\DeclareMathSymbol{\Omega}{\mathalpha}{operators}{10}

\makeatletter
\newcommand{\ifnotempty}[2]{\@ifnotmtarg{#1}{#2}}
\newcommand{\ifempty}[3]{\@ifmtarg{#1}{#2}{3}}
\makeatother

\newcommand{\htarget}[1]{\hypertarget{#1}{#1}}

\newcommand{\superfluid}{\textsc{Superfluid}\xspace}

\newcommand{\mta}[1]{\textsf{#1}}
\newcommand{\axiom}[1]{\textsc{#1}}

\newcommand{\kwd}[1]{\textsf{#1}}
\newcommand{\defkwd}[1]{\textbf{\textsf{#1}}}
\newcommand{\lab}[1]{{\ensuremath{\color{OliveGreen}{\textsf{#1}}}}}
\newcommand{\ctorlab}[1]{{\ensuremath{\color{RawSienna}{\textsf{#1}}}}}
\newcommand{\fieldlab}[1]{{\ensuremath{\color{Plum}{\textsf{#1}}}}}
\newcommand{\datalab}[1]{{\ensuremath{\color{BlueViolet}{\textsf{#1}}}}}
\newcommand{\primlab}[1]{{\ensuremath{\color{RedViolet}{\textsf{#1}}}}}

\newcommand{\caselab}[1]{\fieldlab{$\text{case}_{\datalab{#1}}$}}
\newcommand{\elimlab}[1]{\fieldlab{$\text{elim}_{\datalab{#1}}$}}

\newcommand{\letin}[3]{\kwd{let}\; {#1} = {#2}\; \kwd{in}\; {#3}}
\newcommand{\wildp}{\_}
\newcommand{\univ}{\mathcal{U}}

\newcommand{\code}[1]{\mta{code}\ifnotempty{#1}{\ {#1}}}
\newcommand{\El}[1]{\mta{El}\ifnotempty{#1}{\ {#1}}}
\newcommand{\Code}[1]{\mta{code}\ifnotempty{#1}{\ {#1}}}

\newcommand{\Repr}[1]{\kwd{Repr}\ifnotempty{#1}{\ {#1}}}
\newcommand{\repr}[1]{\kwd{repr}\ifnotempty{#1}{\ {#1}}}
\newcommand{\unrepr}[1]{\kwd{unrepr}\ifnotempty{#1}{\ {#1}}}
\newcommand{\MRepr}{\mta{Repr}}
\newcommand{\Mrepr}{\mta{repr}}
\newcommand{\Munrepr}{\mta{unrepr}}

\newcommand{\Sbody}[1]{\ \left\{\  \begin{aligned}#1\end{aligned} \ \right\}}
\newcommand{\Sdata}[3]{\defkwd{data} \ \datalab{#1} \ifnotempty{#2}{\ {#2}} \ifnotempty{#3}{: {#3} \to \univ}}

\newcommand{\as}{\defkwd{as}}
\newcommand{\by}{\defkwd{by}}
\newcommand{\Sreprvar}[2]{\defkwd{repr} \ {#1}\ \defkwd{as}\ {#2}}

\newcommand{\Ty}{\mta{Ty}}
\newcommand{\Tm}{\mta{Tm}}

\newcommand\tel{\ \textsf{tel}}
\newcommand\type{\ \textsf{type}}
\newcommand\sig{\ \textsf{sig}\ }
\newcommand\op{\ \textsf{op}\ }
\newcommand\IN{^{\textsf{in}}}
\newcommand\DI{^{\textsf{dispIn}}}
\newcommand\D{^{\textsf{dispAlg}}}
\newcommand\COH{^{\textsf{coh}}}
\newcommand\OUT{^{\textsf{out}}}
\newcommand\DO{^{\textsf{dispOut}}}

\newcommand\IND{^{\textsf{ind}}}
\newcommand\INDA{^{\textsf{indAlg}}}
\newcommand\A{^{\textsf{alg}}}
\newcommand\n{\textit}
\newcommand\external{_{\textsf{ext}}}
\newcommand\internal{_{\textsf{int}}}
\newcommand{\Mdata}{\mta{data}}
\newcommand{\Mctor}{\mta{ctor}}
\newcommand{\Melim}{\mta{elim}}

\newcommand{\lang}{\textsc}
\newcommand{\langdata}{{\lang{datatt}}}
\newcommand{\langmltt}{{\lang{mltt}}}
\newcommand{\lambdadata}{\ensuremath{{\lang{datatt}}}\xspace}
\newcommand{\lambdaprog}{\ensuremath{{\lang{prog}}}\xspace}
\newcommand{\lambdamltt}{\ensuremath{{\lang{mltt}}}\xspace}
\newcommand{\R}{\mathcal{R}}
\newcommand{\apR}{\R_{\approx}}

\newenvironment{block}{%
  \par
  \vspace{\abovedisplayskip}
  \centering
  \begingroup
}{%
  \endgroup
  \vspace{\belowdisplayskip}
  \par
}

\newenvironment{definitions}{%
  \par
  \vspace{\abovedisplayskip}
  \centering
  \begingroup
  \setlength{\tabcolsep}{10pt}
  \tabulary{\textwidth}{lL}
}{%
  \endtabulary%
  \endgroup%
  \vspace{\belowdisplayskip}%
  \par\noindent\ignorespaces\hspace{-0.92ex}%
}

\begin{document}

\title{Custom Representations of Inductive Families}
\titlerunning{Custom Representations of Inductive Families}
\author{Constantine Theocharis \orcidID{0009-0001-0198-2750}\and \\ Edwin Brady \orcidID{0000-0002-9734-367X}}
\institute{%
  University of St Andrews, UK \\
  \email{\{kt81,ecb10\}@st-andrews.ac.uk}
}

\maketitle

\begin{abstract}
  Inductive families provide a convenient way of programming with dependent
  types. Yet, when it comes to compilation, their default linked-tree runtime
  representations, as well as the need to convert between different indexed
  views of the same data, can lead to
  unsatisfactory runtime performance. In this paper, we introduce a
  language with dependent types, and inductive families with customisable
  representations. Representations are a version of Wadler's views
  \cite{Wadler1987-zp}, refined to inductive families like in Epigram
  \cite{Mcbride2004-fd}, but with compilation guarantees: a represented
  inductive family will not leave any runtime traces behind, without relying on
  heuristics such as deforestation. This way, we can build a library of
  convenient inductive families based on a minimal set of primitives, whose
  re-indexing and conversion functions are erased during compilation. We show how
  we can express optimisation techniques such as representing
  \datalab{Nat}-like types as GMP-style \cite{gmp} big integers, without special casing in
  the compiler. With dependent types, reasoning about data representations is
  also possible through a provided modality. This yields
  computationally irrelevant isomorphisms between the original and
  represented data.
\end{abstract}


\section{Introduction}\label{sec:intro}

Inductive families are a generalisation of inductive data types found in
programming languages with dependent types. An inductive definition is equipped
with an eliminator that and enables structural recursion over the data, and
captures the notion of mathematical induction. This is a powerful tool for
programming as well as theorem proving. However, this abstraction has a cost
when it comes to compilation: the standard runtime representation of inductive
types is a linked tree structure. This representation is not always the most
efficient and often
forces users to rely on machine primitives to achieve desirable
performance, at the cost of structural induction and dependent pattern matching.

Despite advances in the erasure of irrelevant indices in inductive families
\cite{Brady2004-ay} and the use of theories with irrelevant fragments
\cite{Atkey2018-pj,Moon2021-eb}, there is still a need to convert
between differently-indexed versions of the same data. For example, consider the function that
converts from $\datalab{BinTreeOfHeight}\ T\ n$ to $\datalab{BinTree}\ T$ by forgetting the
height index $n$. This is \emph{not} erased by any current language with dependent
types, unless sized binary trees are defined as a refinement of binary trees
with an erased height field, which hinders dependent pattern matching due to the
presence of non-structural witnesses.

Wadler's views \cite{Wadler1987-zp} provide a way to abstract over inductive
interfaces, so that different views of the same data can be defined and
converted between seamlessly. In the context of inductive families, views have
been used in Epigram \cite{Mcbride2004-fd} that use the index refinement
machinery of dependent pattern matching to avoid certain proof obligations with
eliminator-like constructs. While Wadler's views exhibit a nice way to transport
across a bijection between the original data and the viewed data, they do not
\emph{erase} the view from the final program.

In this paper, we propose an extension $\lambdadata$ to Martin-L\"of type theory \cite{Martin-Lof1984-pz} with
which allows programmers to define inductive types with custom,
correct-by-construction data representations. This is done through user-defined
translations of the constructors and eliminators of an inductive type to a
concrete implementation, which form a bijective view of the original data called
a `representation'. Representations are defined internally to the language, and
require coherence properties that ensure a representation is faithful to its the
original inductive family. We contribute the following:
\begin{itemize}
    \item A formulation of common optimisations such as the `Nat-hack', and
        similarly for other inductive types, as well as zero-cost data reuse when
        reindexing, using custom \emph{representations} (\cref{sec:examples}).
        \item A dependent type system $\lambdadata$ with data types formulated
        in terms of inductive algebras for signatures, along with a translation to
        $\lambdamltt$ that replaces all data types with their defined inductive
        algebras (\cref{sec:type-system}). We have formalised this in Agda (\cref{app:formalisation}).
        \item A prototype implementation of this system in \textsc{Superfluid},
        a programming language with inductive families (\cref{sec:implementation}).
\end{itemize}

\section{A tour of data representations}\label{sec:examples}

A common optimisation done by programming languages with dependent types such as
Idris 2 \cite{idris}, Agda \cite{agda}, Rocq \cite{rocq} and Lean \cite{lean} is
to represent natural numbers more efficiently. The
definition of natural numbers is
\begin{equation}\label{eq:nat-def}
  \Sdata{Nat}{}{} \Sbody{
    \ctorlab{zero} & : \datalab{Nat} \\
    \ctorlab{succ} & : \datalab{Nat} \to \datalab{Nat}
  }
\end{equation}
and generates a case analysis principle $\caselab{Nat}$ of type
\[
   (P : \datalab{Nat} \to \univ)
  \to P\ \ctorlab{zero} \to ((n : \datalab{Nat}) \to P\ (\ctorlab{succ}\ n))
  \to (s : \datalab{Nat}) \to P\ s \,,
\]
which powers pattern matching.
This is a special case of the induction principle $\elimlab{Nat}$, where the inductive hypotheses
are given in each method.
Without further intervention, $\datalab{Nat}$ is represented in unary,
where each digit becomes a heap cell at runtime. This is
inefficient for many basic operations on natural numbers, especially
since computers are well-equipped to deal with numbers natively, so
many real-world implementations will treat \datalab{Nat} specially, swapping the
default inductive type representation with one based on GMP \cite{gmp} integers. This is
done with the replacements
\begin{align*} \label{eq:nat-repr-untyped}
  &|\ctorlab{zero}| = \texttt{0}, \\
  &|\ctorlab{succ}| = \texttt{\textbackslash x => x + 1}, \\
  &|\caselab{Nat}\ P\ m_{\ctorlab{zero}}\ m_{\ctorlab{succ}}\ s| \\
  & \qquad =\ \texttt{let s = $|s|$ in if s == 0 } \\
  &\qquad \qquad \texttt{then $|m_{\ctorlab{zero}}|$} \\
  &\qquad \qquad \texttt{else $|m_{\ctorlab{succ}}|$\ (s - 1)},
\end{align*}
where $|\cdot|$ denotes a source translation into a compilation target
language with appropriate big unsigned integer primitives. This is the `Nat-hack'.\footnote{
Idris 2 will in fact look for any \datalab{Nat}-like types and apply this optimisation.
A similar optimisation is also done with list-like and boolean-like types because
they have a canonical representation in the target runtime, Chez Scheme.}

In addition to the constructors and case analysis, the compiler might define
translations for commonly used definitions which have a more efficient
counterpart in the target, such as addition, multiplication,
etc. The recursively-defined functions are well-suited to proofs and reasoning,
while the primitives are more efficient for computation. This way, the surface
program can take advantage of the structural properties of induction,
while still benefiting from an efficient representation at runtime.

Unfortunately, this approach only works for the data types which the
compiler recognises as `special'. Particularly in the presence of dependent
types, other data types might end up being equivalent to \datalab{Nat} or
another `nicely-representable' type, but in a non-trivial way that the compiler
cannot recognise. It is also hard to know when such optimisations will fire
and performance can become brittle upon refactoring.
Hence, one of our goals is to extend this optimisation to work
for any data type. To achieve this, our framework requires that
representations are fully typed in a way that ensures the behaviour of the
representation of a data type matches the behaviour of the data type itself.

\subsection{The well-typed Nat-hack}

A representation definition looks like
\[
  \Sreprvar{\datalab{Nat}}{\primlab{UBig}} \Sbody{
    \ctorlab{zero}\ \as&\ \primlab{0} \\
    \ctorlab{succ}\ n\ \as&\ \primlab{1}\,\primlab{$+$}\,n \\
    \elimlab{Nat}\ \as&\ \primlab{ubig-elim}\\ \by&\ \primlab{ubig-elim-zero-id}, \\ &\ \primlab{ubig-elim-add-one-id}
  }
\]
The inductive type $\datalab{Nat}$ is represented as the type $\primlab{UBig}$ of big
unsigned integers, with translations for the constructors
$\ctorlab{zero}$ and $\ctorlab{succ}$, and the eliminator $\elimlab{Nat}$ (now with the inductive hypotheses).
Additionally, the eliminator must satisfy the computation rules of the
$\datalab{Nat}$ eliminator, which are postulated as propositional equalities.
This representation is valid in a context containing the symbols
\begin{gather*}
  \primlab{0}, \primlab{1} : \primlab{UBig} \qquad
  \primlab{$+$} : \primlab{UBig} \to \primlab{UBig} \to \primlab{UBig} \\
  \begin{aligned}
  \primlab{ubig-elim} :\ &(P : \primlab{UBig} \to \univ) \to P\ \primlab{0}
  \to ((n : \primlab{UBig}) \to \overline{P\ n} \to P\ (\primlab{1}\,\primlab{$+$}\,n)) \\
   &\to (s : \primlab{UBig}) \to P\ s
  \end{aligned}
\end{gather*}
and propositional equalities
\begin{align*}
  &\primlab{ubig-elim-zero-id} : _{\forall Pbr}  \primlab{ubig-elim}\ P\ b\ r\ \primlab{0} = b \\
  &\primlab{ubig-elim-add-one-id} : _{\forall Pbrn}  \primlab{ubig-elim}\ P\ b\ r\ (\primlab{1}\,\primlab{$+$}\,n)
  = r\ n\ (\lambda \wildp. \ \primlab{ubig-elim}\ P\ b\ r\ n)  \,.
\end{align*}
For the remainder of the paper we will work with eliminators rather than case
analysis but the approach can be specialised to the latter if the language has
general recursion. Assuming call-by-value semantics, inductive hypotheses are labelled
$\overline{P}$ which denotes lazy values, that is, functions $\datalab{Unit} \to
P$.

The compiler knows how to perform pattern matching on $\datalab{Nat}$, and
produces invocations of $\elimlab{Nat}$ as a result. On the other hand, it does
not know how to pattern match on $\primlab{UBig}$. With this representation, we
get the best of both worlds: \datalab{Nat} is used for pattern matching, but is
then replaced with $\primlab{UBig}$ during compilation, generating more efficient code. We
expect that the underlying implementation of $\primlab{UBig}$ indeed satisfies
these postulated properties, which is a separate concern from the correctness of
the representation itself. However, such postulates are only needed when the
representation target is a primitive; the next examples use defined types as
targets, so that the coherence of the target eliminator follows from the
coherence of other eliminators used in its implementation.

\subsection{Vectors as a refinement of lists}

In addition to representing inductive types as primitives, we can use
representations to share the underlying data when converting between indexed
views of the same data. For example, we can define a representation of
$\datalab{Vec}$ in terms of $\datalab{List}$, so that the conversion from one to
the other is `compiled away'. We can do this by representing the indexed type as
a refinement of the unindexed type by an appropriate relation. For the case of
$\datalab{Vec}$, we know intuitively that
\[
  \datalab{Vec}\ T\ n \simeq \{ l : \datalab{List}\ T \mid \lab{length}\ l = n \}
\]
which we shorthand as $\lab{List'}\ T\ n := \{ l : \datalab{List}\ T \mid
\lab{length}\ l = n \}$. We will take the subset $\{ x : A \mid P\ x \}$ to mean
a $\Sigma$-type $(x : A) \times P\ x$ where the right component is irrelevant
and erased at runtime. We also assume that $\datalab{Vec}$'s $n$
index is computationally irrelevant.
We can thus choose $\lab{List'}\ T\ n$ as the representation of
$\datalab{Vec}\ T\ n$. We are then tasked with providing terms that correspond to
the constructors of $\datalab{Vec}$ but for $\lab{List'}$. These can be defined
as
\[
  \begin{aligned}
  & \lab{nil} : \lab{List'}\ T\ \ctorlab{zero} \\
  & \lab{nil} = (\ctorlab{nil}, \ctorlab{refl})
  \end{aligned}\qquad
  \begin{aligned}
  & \lab{cons} : T \to \lab{List'}\ T\ n \to \lab{List'}\ T\ (\ctorlab{succ}\ n) \\
  & \lab{cons}\ x\ (\mathit{xs}, p) = (\ctorlab{cons}\ x\ \mathit{xs}, \lab{cong}\ (\ctorlab{succ})\ p)
  \end{aligned}
\]
Next we need to define the eliminator for $\lab{List'}$, which should have the form
\begin{align*}
  & \lab{elim-List'} : (E : (n : \datalab{Nat}) \to \lab{List'}\ T\ n \to \lab{Type}) \\
  & \quad \to E\ \ctorlab{zero}\ \lab{nil} \\
  & \quad \to ((x : T) \to (n : \datalab{Nat}) \to (\mathit{xs} : \lab{List'}\ T\ n) \to \overline{E\ n\ xs} \to E\ (\ctorlab{succ}\ n)\ (\lab{cons}\ x\ \mathit{xs})) \\
  & \quad \to (n : \datalab{Nat}) \to (v : \lab{List'}\ T\ n) \to E\ n\ v
\end{align*}
Dependent pattern matching does a lot of the heavy lifting by refining the
length index and equality proof by matching on the underlying list. However we still need to
substitute the lemma $\lab{cong}\ (\ctorlab{succ})\ (\lab{\ctorlab{succ}-inj}\ p) = p$ in the recursive case.
\begin{align*}
  &\lab{elim-List'}\ P\ b\ r\ \ctorlab{zero}\ (\ctorlab{nil}, \ctorlab{refl}) = b \\
  &\lab{elim-List'}\ P\ b\ r\ (\ctorlab{succ}\ m)\ (\ctorlab{cons}\ x\ \mathit{xs}, e) = \lab{subst}\
  \begin{aligned}[t]
  &(\lambda p .\ P\ (\ctorlab{succ}\ m)\ (\ctorlab{cons}\ x\ \mathit{xs}, p))\ \\
  &\hspace{-3em} (\lab{\ctorlab{succ}-cong-id}\ e)\ (r\ x\ ({\mathit{xs}}, \lab{\ctorlab{succ}-inj}\ e) \\
  &\hspace{-3em}  (\lambda \wildp . \ \lab{elim-List'}\ P\ b\ r\ m\ ({\mathit{xs}}, \lab{\ctorlab{succ}-inj}\ e)))
  \end{aligned}
\end{align*}
Finally, we need to prove that the eliminator satisfies the expected computation
rules propositionally. These are
\begin{align*}
  \lab{elim-List'-nil-id} : &\ \lab{elim-List'}\ P\ b\ r\ \ctorlab{zero}\ (\ctorlab{nil}, \ctorlab{refl}) = b \\[0.5em]
  \lab{elim-List'-cons-id} :&\  \lab{elim-List'}\ P\ b\ r\ (\ctorlab{succ}\ m)\ (\ctorlab{cons}\ x\ \mathit{xs}, \lab{cong}\ \ctorlab{succ}\ p) \\
  & = r\ x\ (\mathit{xs}, p)\ (\lambda \wildp .\ \lab{elim-List'}\ P\ b\ r\ m\ (\mathit{xs}, p))
\end{align*}
The first holds definitionally, and the second requires a small amount of
equality reasoning.
This completes the definition of the representation of
$\datalab{Vec}$ as $\lab{List'}$, which would be written as
\[
  \Sreprvar{\datalab{Vec}\ T\ n}{\lab{List'}\ T\ n} \Sbody{
    \ctorlab{nil}\ \as&\ \lab{nil} \\
    \ctorlab{cons}\ \as&\ \lab{cons} \\
    \elimlab{Vec}\ \as&\ \lab{elim-List'} \\
     \by&\ \lab{elim-List'-nil-id}, \\ &\ \lab{elim-List'-cons-id}
  }
\]
Now the hard work is done. Every time we are working with a $v : \datalab{Vec}\
T\ n$, its form will be $(l, p)$ at runtime, where $l$ is the underlying list
and $p$ is the proof that the length of $l$ is $n$. Under the assumption that
the $\Sigma$-type's right component is irrelevant and erased at runtime, every
vector is simply a list at runtime, where the length proof has been erased. In
practice, this erasure is achieved in \superfluid using quantitative type theory
\cite{Atkey2018-pj}. In \cref{sub:irr} we show how to formally identify
computationally irrelevant conversion functions.

We can utilise this representation to convert between $\datalab{Vec}$ and
$\datalab{List}$ at zero runtime cost. We can do this using the $\repr{}$ and $\unrepr{}$
operators of the language (defined in \cref{sec:type-system}). These allow us to convert
between an inductive type and its representation. Specifically, we
can define the functions
\begin{align*}
  &\lab{forget-length} : \datalab{Vec}\ T\ n \to \datalab{List}\ T \\
  &\lab{forget-length}\ v = \letin{(l, \wildp)}{\repr{v}}{l} \\[1em]
  &\lab{remember-length} : (l : \datalab{List}\ T) \to \datalab{Vec}\ T\ (\lab{length}\ l) \\
  &\lab{remember-length}\ l = \unrepr{(l, \ctorlab{refl})} \,.
\end{align*}
In \cref{sub:irr} we will show that such functions are inverses
of one another and are computationally irrelevant.
These operators are typed as
\[
  \Mrepr : A \to \MRepr\ A \qquad \Munrepr : \MRepr\ A \to A
\]
where $\MRepr\ A$ computes to the defined representation of $A$, if $A$ is a
data type. $\MRepr$ is a kind of `intensional' modality, with the property that
$\MRepr\ A \simeq A$. It allows us to transport across equivalences introduced by
representations in a computationally-irrelevant manner.

\subsection{General reindexing}

The idea from the previous example can be generalised to any data type. In
general, suppose that we have two inductive families
\[
 \datalab{F} : P \to \univ \qquad \datalab{G} : (p : P) \to X\ p \to \univ
\]
for some index family $X : P \to \univ$. If we hope to represent $\datalab{G}$
as some refinement of $\datalab{F}$ then we must provide a way to
compute $\datalab{G}$'s extra indices $X$ from $\datalab{F}$, like we computed
$\datalab{Vec}$'s extra $\datalab{Nat}$ index from $\datalab{List}$ with
$\lab{length}$ in the previous example. This means that we need to provide a
function $\lab{comp} : {\forall p.}\ \datalab{F}\ p \to X\ p$ which can then be
used to form the family
\[
  \datalab{F}^{\lab{comp}}\ p\ x :=  \{ f : \datalab{F}\ p \mid \lab{comp}\ f = x \} .
\]
If $\datalab{G}$ is `equivalent' to the algebraic ornament of $\datalab{F}$ by
the algebra defining $\lab{comp}$ (given by an isomorphism between the
underlying polynomial functors), then it is also equivalent to the $\Sigma$-type
above. The `recomputation lemma' of algebraic ornaments \cite{Dagand2012-aw}
then arises from its projections. Our system allows us to \emph{set} the
representation of $\datalab{G}$ as $\datalab{F}^{\lab{comp}}$, so that the
forgetful map from $\datalab{G}$ to $\datalab{F}$ as well as the recomputation
map from $\datalab{F}$ to $\datalab{G}$ are erased, constructed
with $\Mrepr$ and $\Munrepr$. We formulate this pattern in the general setting in \cref{sub:irr}.

\subsection{Zero-copy deserialisation}

The machinery of representations can be used to implement zero-copy deserialisation
of data formats into inductive types. Here we sketch how this could work. Consider the following
record for a player in a game:
\[
  \Sdata{Player}{}{} \Sbody{
    \ctorlab{player} : &\ (\fieldlab{position} : \datalab{Position}) \\
    \to &\ (\fieldlab{direction} : \datalab{Direction}) \\
    \to &\ (\fieldlab{items} : \datalab{Fin}\ \lab{MAX\_INVENTORY}) \\
    \to &\ (\fieldlab{inventory} : \datalab{Inventory}\ (\lab{fin-to-nat}\ \fieldlab{items})) \to \datalab{Player}
  }
\]
We can use the $\datalab{Fin}$ type to maintain the invariant that the inventory
has a maximum size. Additionally, we can index the $\datalab{Inventory}$ type by
the number of items it contains, which might be defined similarly to $\datalab{Vec}$:
\[
  \Sdata{Inventory}{(n : \datalab{Nat})}{} \Sbody{
    \ctorlab{empty} : &\ \datalab{Inventory}\ \ctorlab{zero} \\
    \ctorlab{add} : &\ \datalab{Item} \to \datalab{Inventory}\ n \to  \datalab{Inventory}\ (\ctorlab{succ}\ n)
  }
\]
We can use the full power of inductive families to model the domain of our
problem in the way that is most convenient for us. If we were writing this in a
lower-level language, we might choose to use the serialised format directly when
manipulating the data, relying on the appropriate pointer arithmetic to access
the fields of the serialised data, to avoid copying overhead. Representations
allow us to do this while still being able to work with the high-level inductive
type.

We can define a representation for $\datalab{Player}$
as a pair of a byte buffer and a proof that the byte buffer contents correspond to
a player record. Similarly, we can define a representation for $\datalab{Inventory}$
as a pair of a byte buffer and a proof that the byte buffer contents correspond to
an inventory record of a certain size. By the implementation of the eliminator
in the representation, the projection $\fieldlab{inventory} : (p :
\datalab{Player}) \to \datalab{Inventory}\ p.\fieldlab{items}$ is compiled into
some code to slice into the inventory part of the player's byte buffer. We
assume that the standard library already represents $\datalab{Fin}$ in the same
way as $\datalab{Nat}$, so that reading the $\fieldlab{items}$ field is a
constant-time operation (we do not need to build a unary numeral). We can thus
define the representation
of $\datalab{Player}$ as
\[
  \Sreprvar{\datalab{Player}}{\{ \datalab{Buf} \mid \datalab{IsPlayer} \}} \Sbody{
    \ctorlab{player}\ \as&\ \lab{buf-is-player} \\
    \elimlab{Player}\ \as&\ \lab{elim-buf-is-player} \\
                      \by&\ \lab{elim-buf-is-player-id}
  }
\]
with an appropriate definition of $\datalab{IsPlayer}$ which refines a byte
buffer. The refinement would have to match the expected structure of the byte
buffer, so that all the required fields can be extracted. Allais
\cite{Allais2023-zq} explores how data descriptions that index into a flat
buffer can be defined.

\subsection{Transitivity}\label{sub:transitivity}

Representations are transitive, so in the previous example, the eventual
representation of $\datalab{Vec}$ at runtime is determined by the representation
of $\datalab{List}$. It is possible to define a custom representation for
$\datalab{List}$ itself, for example a heap-backed array or a finger tree, and
$\datalab{Vec}$ would inherit this representation. However it will still be the
case that $\Repr{(\datalab{Vec}\ T\ n)} \equiv \datalab{List}\ T$, which means
the $\Repr{}$ modality only considers the immediate defined
representation of a term. Regardless, we can construct predicates that find
types which satisfy a certain eventual representation. For example, given a
\datalab{Buf} type of byte buffers, we can consider the set of all types which
are eventually
represented as a \datalab{Buf}:
\[
  \Sdata {ReprBuf}{(T : \univ)}{} \Sbody{
    \ctorlab{buf} &: \datalab{ReprBuf}\ \datalab{Buf} \\
    \ctorlab{from} &: \datalab{ReprBuf}\ (\Repr{T}) \to \datalab{ReprBuf}\ T \\
    \ctorlab{refined} &: \datalab{ReprBuf}\ T \to \datalab{ReprBuf}\ \{ t : T \mid  P\ t \}
  }
\]
Every such type comes with a projection function to the \datalab{Buf} type
\begin{align*}
  &\lab{as-buf} :\{r : \datalab{ReprBuf}\ T\} \to T \to \datalab{Buf} \\
  &\lab{as-buf}\ \{r = \ctorlab{buf}\}\ x = x \\
  &\lab{as-buf}\ \{r = \ctorlab{from}\ t\}\ x = \lab{as-buf}\ t\ (\repr{x}) \\
  &\lab{as-buf}\ \{r = \ctorlab{refined}\ t\}\ (x, \wildp) = \lab{as-buf}\ t\ x
\end{align*}
which eventually computes to the identity function after applying $\repr{}$ the
appropriate amount of times. Upon compilation, every type is converted to its
eventual representation, and all $\repr{}$ calls are erased, so the
$\lab{as-buf}$ function becomes the identity function at
runtime, given that the $r$ argument is known at compile-time and monomorphised.

\section{A type system for data representations}\label{sec:type-system}

In this section, we develop an extension of dependent type theory with inductive
families and custom data representations. We start in \cref{sub:algebras} by
exploring the semantics of data representations in terms of inductive algebras
for signatures. In \cref{sub:lambdadata} we define a core language $\lambdadata$
with these features. The base theory is intensional Martin-L\"of type theory
(\lambdamltt) \cite{Martin-Lof1984-pz} with a single universe $\univ : \univ$.
We omit considerations of consistency and universe hierarchy, though these can
be added if needed. In \cref{sub:lambdadata}, we define the modality $\MRepr$
that allows us to convert between inductive types and their representations.
Finally, in \cref{sub:translation} we define a translation from $\lambdadata$ to
\emph{extensional} \lambdamltt, which `elaborates away' all inductive families
to their representations. All of the examples in the paper so far have been
written in a surface language that elaborates to $\lambdadata$.

The languages we work with are defined in an intrinsically well-formed manner \cite{Altenkirch2016-zc}
as a setoid over definitional equality, with de-Bruijn
indices for variables. Weakening of terms is generally left implicit to reduce
syntactic noise, and often named notation is used when indices are
implied. We use $(a : A) \to B$ for dependent functions, $(a : A) \times B$ for
dependent pairs, $a \equiv_A a'$ for propositional equality, and $a = a' : A$
for definitional equality. Substitution is denoted with square brackets: if
$\Gamma, A \vdash B$ and $\Gamma \vdash a : A$ then $\Gamma \vdash B[a]$. We
also notationally identify elements of the universe $A : \univ$ with types $A
\type$.
Besides the
usual judgement forms of \lambdamltt, we also have telescopic judgement forms
\begin{definitions}
$\Gamma \vdash \Delta \tel$                & $\Delta$ is a telescope in $\Gamma$, \\
$\Gamma \vdash \delta :: \Delta$           & $\delta$ is a spine (list of terms) matching telescope $\Delta$,
\end{definitions}
with accompanying rules shown in \cref{fig:tel-rules}.

\begin{figure}[H]
  \vspace{-1em}
  \begin{mathpar}
  \inferrule[\htarget{Tel-Empty}]{ }{\Gamma \vdash \bullet \tel}
  \and
  \inferrule[\htarget{Tel-Extend}]{\Gamma \vdash A \type \\ \Gamma, A \vdash \Delta \tel}{\Gamma \vdash (A, \Delta) \tel}
  \\
  \inferrule[\htarget{Spine-Empty}]{ }{\Gamma \vdash (\,) :: \bullet} \and
  \inferrule[\htarget{Spine-Extend}]{\Gamma \vdash a : A \\ \Gamma \vdash \delta : \Delta[a]}{\Gamma \vdash (a, \delta) :: (A, \Delta)}
  \vspace{-1em}
  \end{mathpar}
  \caption{Rules for forming telescopes and spines.}
  \label{fig:tel-rules}
  \vspace{-1em}
\end{figure}

Extending contexts by telescopes (such as $\Gamma, \Delta$) is defined by
induction on telescopes. We write $\Delta \to X$ for an iterated function type
with codomain $\Gamma, \Delta \vdash X$, and $(\delta :: \Delta) \to X[\delta]$
when names are highlighted. We will often use the notation $\delta.y$ to extract
a certain index $y$ from a spine $\delta$. This is used when we define
telescopes using named notation. For example, if $\delta :: (\textit{X} : A \to
\univ, \textit{y} : (a : A) \to X\ a)$, then $\delta.X : A \to \univ$ and
$\delta.y : (a : A) \to \delta.X\ a$.

\newcommand{\ValidCase}{\mta{ValidCase}}

\subsection{Algebraic signatures}

A representation of a data type must be able to emulate the behaviour of the
original data type. In turn, the behaviour of the original data type is
determined by its elimination, or induction principle. This means that a
representation of a data type should provide an implementation of induction of
the same `shape' as the original. Induction can be characterised in terms of
algebras and displayed algebras of algebraic signatures \cite{Adamek2010-ls,Kovacs2023-gq}.
Algebraic signatures consist of a list of operations, each with a specified
arity. There are many flavours of algebraic signatures with varying degrees of
expressiveness. For this paper, we are interested in the ones which can
be used as a syntax for defining inductive families in a type theory. Thus, we define
two new judgement forms
\begin{definitions}
$\Gamma \vdash S \sig \Delta$    & $S$ is a signature with indices $\Delta$ in context $\Gamma$ \\
$\Gamma \vdash O \op \Delta$     & $O$ is an operation with indices $\Delta$ in context $\Gamma$ ,
\end{definitions}
with accompanying rules shown in \cref{fig:tel-rules}.
Signatures are lists of operations, and operations build up constructor types.

\enlargethispage{\baselineskip}

\begin{figure}[H]
  \begin{mathpar}
    \inferrule[\htarget{Sig-Empty}]{\Gamma \vdash \Delta \tel}{\Gamma \vdash \epsilon \sig \Delta}
    \and
    \inferrule[\htarget{Sig-Extend}]{\Gamma \vdash \Delta \tel \\ \Gamma \vdash O \op \Delta \\ \Gamma \vdash S \sig \Delta}{\Gamma \vdash (O \lhd S) \sig \Delta}
    \\
    \inferrule[\htarget{Op-Ext}]{\Gamma \vdash A \type \\ \Gamma, A \vdash O \op \Delta}
    {\Gamma \vdash (A\ \to\external O) \op \Delta} \and
    \inferrule[\htarget{Op-Int}]{\Gamma \vdash \delta :: \Delta \\ \Gamma \vdash O \op \Delta}
    {\Gamma \vdash (\iota\ \delta \to\internal O) \op \Delta} \and
    \inferrule[\htarget{Op-Ret}]{\Gamma \vdash \delta :: \Delta}
    {\Gamma \vdash (\iota\ \delta) \op \Delta}
    \end{mathpar}
  \caption{Rules for forming signatures and operations.}
\end{figure}

Each signature is described by an associated telescope of indices $\Delta$, and a
\emph{finite list} of operations:
\begin{itemize}
    \item $(x : A) \to\external O[x]$, a (dependent) abstraction over some external type $A$, of another operation $O$.
    \item $\iota\ \delta \to\internal \ O$, an
      abstraction over a recursive occurence of
      the object being defined, with indices $\delta$, of another operation
      $O$.
    \item $\iota\ \delta$, a constructor of the object being defined, with indices $\delta$.
\end{itemize}

\begin{example}[Natural numbers]\label{ex:nat-sig}
The signature for natural numbers is indexed by the empty telescope $\bullet$.
It is defined as $\Gamma \vdash (\iota\ (\,) \lhd \iota\ (\,) \to\internal
\iota\ (\,) \lhd \epsilon) \sig \bullet$. We can add labels to aid readability,
omit index spines if they are empty, and omit the final $\epsilon$ from
signatures:
\[
\Gamma \vdash (\textit{zero} : \iota \lhd \textit{succ} : \iota\ \to\internal
\iota) \sig \bullet \,.
\]
\end{example}

\begin{example}[Vectors]\label{ex:vec-sig}
The signature for vectors of elements of type $T$ and length $n$ is indexed by
the telescope $(T : \univ, n : \mathbb{N})$, defined as
\begin{align*}
\Gamma \vdash (&\n{nil} : (T : \univ) \to\external \iota\ T\ \n{zero}\ \lhd \\
&\n{cons} : (T : \univ) \to\external (n' : \mathbb{N}) \to\external (t : T)
\to\external \iota\ T\ n' \to\internal \iota\ T\ (\n{succ}\ n')) \\ &\sig (T :
\univ, n : \mathbb{N}) \,.
\end{align*}
Later (\cref{sub:constructing-inductive-families}) we will see how we can use the signature in \cref{ex:nat-sig} to define
the type of natural numbers $\Gamma \vdash \mathbb{N} \type$.
\end{example}

Notice that this syntax only allows occurrences of $\iota$ in positive
positions, which is a requirement for inductive types. Different classes of
algebraic signatures, theories and quantification are explored in detail by
Kov\'acs \cite{Kovacs2023-gq}. We make no distinction between parameters and
indices, though it is possible to add parameters by augmenting the syntax for
signatures with an extra telescope that must be uniform across operations.

\subsection{Interpreting signatures in the type theory} \label{sub:algebras}

In order to make use of our definition for algebraic signatures, we would like
to be able to interpret their structure as types in the type theory we are working
with.

\subsubsection{Algebras}

An \emph{algebra} for a signature $\Gamma \vdash S \sig \Delta$ and carrier type
$\Gamma, \Delta \vdash X \type$ interprets the structure of $S$ in terms of the
type $X$. Concretely, this produces a telescope which matches the structure of
$S$ but replaces each occurrence of $\iota\ \delta$ with $X[\delta]$. The
function arrows $\to\internal$ and $\to\external$ in $S$ are interpreted as the
function arrow $\to$ of the type theory.

\begin{example}[Natural numbers]\label{ex:nat-alg}
An algebra for the signature of natural numbers (\cref{ex:nat-sig}) over a carrier $\Gamma \vdash N \type$ is a spine matching
the telescope \[\Gamma \vdash (\n{zero}: N,\ \n{succ}: N \to N) \tel.\]
\end{example}

\begin{example}[Vectors]\label{ex:vec-alg}
An algebra for the signature of vectors (\cref{ex:vec-alg}) over a carrier $\Gamma, T : \univ, n : \mathbb{N} \vdash V \type$ is a spine matching
the telescope
\begin{align*}
 \Gamma \vdash (&\n{nil} : (T : \univ) \to V[T, \n{zero}], \\
&\n{cons} : (T : \univ) \to (n' : \mathbb{N}) \to (t : T) \to (ts : V[T,n']) \to V[T, \n{succ}\ n']) \tel.
\end{align*}
\end{example}

\subsubsection{Induction}

The actual type of natural numbers $\Gamma \vdash \mathbb{N} \type$ is the
carrier of an algebra over the signature of natural numbers. In particular, the
`best' such algebra: one whose operations do not forget any information. In the
language of category theory, this is the initial algebra in the category of
algebras over the signature of natural numbers. An equivalent formulation of
initial algebras is algebras which support \emph{induction}, which is more
suitable for our (syntactic) purposes. An algebra $\alpha :: (\textit{zero}: X,\
\textit{succ}: X \to X)$ for natural numbers supports induction if:

\begin{quote}
For any type family $X \vdash Y \type$, if we can construct a $\textit{zero}_Y :
Y[\alpha.\textit{zero}]$ and a $\textit{succ}_Y : (x : X) \to Y[x] \to
Y[\alpha.\textit{succ}\ x]$, then we can construct a $\sigma[x] : Y[x]$ for all
$x : X$.
\end{quote}

The type family $Y$ is commonly called the \emph{motive}, and $(\textit{zero}_Y,
\textit{succ}_Y)$ are the \emph{methods}. The produced term family $x : X \vdash
\sigma : Y[x]$ is a \emph{section} of the type family $Y$. Induction also
requires that the section acquires its values from the provided methods. This
means that
\[
\sigma[\alpha.\textit{zero}] = \textit{zero}_Y \qquad \sigma
[\alpha.\textit{succ}\ x] = \textit{succ}_Y\ x\ \sigma[x].
\]
We call these \emph{coherence conditions}. A section that satisfies these
conditions is called a \emph{coherent section}. These equations might or might
not hold definitionally. In the former case, we have the definitional equalities
\begin{align*}
\Gamma &\vdash \sigma[\alpha.\textit{zero}] = \textit{zero}_Y : Y[\alpha.\textit{zero}] \\
\Gamma, x : X &\vdash \sigma[\alpha.\textit{succ}\ x] = \textit{succ}_Y\ x\ \sigma[x] : Y[\alpha.\textit{succ}\ x].
\end{align*}
In the latter case, we have a spine of propositional equality witnesses
\begin{align*}
\Gamma \vdash \sigma_{\text{coh}} :: (&\n{zero}_{\text{coh}} :
\sigma[\alpha.\textit{zero}] \equiv \textit{zero}_Y, \\
&\n{succ}_{\text{coh}} : (x : X) \to \sigma[\alpha.\textit{succ}\ x] \equiv
\textit{succ}_Y\ x\ \sigma[x]).
\end{align*}

\subsubsection{Displayed algebras}

Notice that the methods $(\textit{zero}_Y, \textit{succ}_Y)$ look like an
algebra for the signature of natural numbers too, but their carrier is now a
type family over another algebra carrier $X$, and the types of the operations
mention both $X$ and $Y$, using $\alpha$ to go from $\Delta$ to $X$. These are
\emph{displayed algebras}. In general, a displayed algebra for a signature
$\Gamma \vdash S \sig \Delta$, algebra $\alpha$ for $S$ over carrier $\Gamma,
\Delta \vdash X \type$, and carrier family $\Gamma, \Delta, X \vdash Y \type$,
interprets the structure of $S$ in terms of both $X$ and $Y$. This produces a
telescope which matches the structure of $S$ but replaces each recursive
occurrence $\iota\ \delta$ with an argument $x : X$ as well as an argument $y :
Y[x]$. Each operation returns a $Y$ with indices computed from $\alpha$. Again,
the function arrows in $S$ are interpreted as function types in the type theory.

\begin{example}[Vectors]\label{ex:vec-disp-alg}
A displayed algebra for an algebra $(\n{nil}, \n{cons})$ for vectors
(\cref{ex:vec-alg}) over a carrier family $\Gamma, T : \univ, n : \mathbb{N}, v
: V[T, n] \vdash W \type$ is a spine matching
the telescope
\begin{align*}
 \Gamma \vdash (&\n{nil}_W : (T : \univ) \to W[T, \n{zero}, \n{nil}\ T], \\
&\n{cons}_W : (T : \univ) \to (n' : \mathbb{N}) \to (t : T) \to (ts : V[T,n']) \\
&\qquad \to (ts_W : W[T, n', ts]) \to W[T, \n{succ}\ n', \n{cons}\ T\ n'\ t\ ts]) \tel.
\end{align*}
\end{example}

In practice, in a call-by-value setting, it is desirable for the inductive
hypotheses of a displayed algebra ($\n{ts}_W$ above) to be
\emph{lazy} values. This improves performance when the inductive hypotheses are
not needed. We leave this as an implementation detail.

Finally, we come to the central definition
that classifies the algebras which
support induction:
\begin{definition}
An algebra is \emph{inductive} if every displayed algebra over it has a coherent section.
\end{definition}
The elimination rule for inductive data types in programming languages is
exactly this: given any motive and methods (a displayed algebra), we get a
dependent function from the type of the scrutinee to the type of the motive (a
section). Furthermore this function satisfies some appropriate computation rules:
when we plug in a constructor, we get the result of the method corresponding to it (the coherence conditions).
Usually in programming languages, these conditions hold definitionally, as they are the primary means
of computation with data.




\subsection{Defining algebras and friends}

In order to utilise these constructions for our type system, we now explicitly define
the follwing objects:
\begin{definitions}
$\Gamma \vdash S\A\ X \tel$ & Algebras for a signature $\Gamma \vdash S \sig \Delta$ over a carrier $\Gamma, \Delta \vdash X \type$. \\
$\Gamma \vdash \alpha\D\ Y \tel$ & Displayed algebras for an algebra $\Gamma \vdash \alpha :: S\A\ X$ over a motive $\Gamma, \Delta, X \vdash Y \type$. \\
$\Gamma \vdash \beta\COH\ \sigma \tel$ & Propositional coherence for a section $\Gamma, \Delta, X \vdash \sigma : Y$ of a displayed algebra $\Gamma \vdash \beta :: \alpha\D\ Y$.
\end{definitions}
All constructions labelled with superscripts are not part of the syntax of the
type system, but rather functions in the metatheory which compute syntactic
objects such as telescopes.

The algebras for a signature are defined
by case analysis on $S$:
\begin{align*}
&\boxed{\Gamma \vdash S\A\ X \tel} \\
&\epsilon\A\ X = \bullet \qquad (O \lhd S')\A\ X = ((\nu :: O\IN\ X) \to X[\nu\OUT],\ {S'}\A\ X) \,.
\end{align*}
An empty signature $\epsilon$ produces an empty telescope, while an extended
signature $O \lhd S'$ produces a telescope extended with a function
corresponding to $O$. This function goes from the inputs of $O$ interpreted in
$X$, to $X$ evaluated at the output indices.
The inputs and outputs of each operation $O$ in an algebra are defined by case
analysis on $O$:
\begin{block}
\setlength{\tabcolsep}{10pt}
\begin{tabular}{ll}
$\boxed{\Gamma \vdash O\IN\ X \tel}$ & \boxed{\Gamma \vdash \{O\}\ \nu\OUT :: \Delta} \\[1em]
{$\begin{aligned}
&(A\to\external O')\IN\ X = (a : A,\ O'[a]\IN\ X) \\
&(\iota\ \delta \to\internal O')\IN\ X = (x : X[\delta],\ {O'}\IN\ X) \\
&(\iota\ \delta)\IN\ X = \bullet
\end{aligned}$}
&
{$\begin{aligned}
&\{O = A \to\external O'\}\ (a, \nu')\OUT = {\nu'}\OUT \\
&\{O = \iota\ \delta \to\internal O'\}\ (x,\ \nu')\OUT = {\nu'}\OUT  \\
&\{O = \iota\ \delta\}\ (\,)\OUT = \delta \,.
\end{aligned}$}
\end{tabular}
\end{block}

A similar construction can be performed for displayed algebras over algebras.
Displayed algebras are defined by case analysis on $S$, which is an implicit parameter of $-\D$:
\begin{align*}
&\boxed{\Gamma \vdash \{S\}\ \alpha\D\ Y \tel} \\
& \{S = \epsilon\}\ (\,)\D\ Y = \bullet \\
&\{S = O \lhd S'\}\ (\alpha_O, \alpha')\D\ Y = ((\mu :: \alpha_O\DI\ Y) \to Y[\mu\DO],\ {\alpha'}\D\ Y) \,.
\end{align*}
We omit the definitions of $-\DI$ and $-\DO$ and
\href{https://github.com/kontheocharis/rep-agda/blob/e6bd34adaab630f5787c63a95fa86869f6c19da4/TT/Sig.agda\#L107}{refer to the Agda formalisation},
but they are similar to
the definitions of $-\IN$ and $-\OUT$.

Next we define the coherence conditions for a displayed algebra as a telescope
\begin{align*}
&\boxed{\Gamma \vdash \{S\}\ \{\alpha\}\ \beta\COH\ \sigma \tel} \\
& \{S = \epsilon\}\ \{\alpha = (\,)\}\ (\,)\COH\ \sigma = \bullet \\
& \{S = O \lhd S'\}\ \{\alpha = (\alpha_O, \alpha')\}\ (\beta_O, \beta')\COH\ \sigma \\
&\qquad = ((\nu :: O\IN\ X) \to \sigma[\alpha_O\ \nu] \equiv \beta_O\ (\sigma \, \$ \, \nu),\  {\beta'}\COH\ \sigma) \,.
\end{align*}
The notation $\sigma \, \$\, \nu$ applies the section $\sigma$ to the input $\nu$,
yielding a displayed input by sampling the section to get the inductive
hypotheses.

Now we can define induction for an algebra $\alpha$ as a type
\begin{align*}
&\boxed{\Gamma \vdash \{S\}\ \alpha\IND \type} \\
&\{S\}\ \alpha\IND = (Y : (\delta :: \Delta) \to X[\delta] \to \univ) \to (\beta :: \alpha\D\ (\delta.\ x.\ Y\ \delta\ x)) \\
&\qquad\to (\sigma : (\delta :: \Delta) \to (x : X[\delta]) \to Y\ \delta\ x)) \times (\rho :: \beta\COH\ (\delta.\ x.\ \sigma\ \delta\ x)) \,,
\end{align*}
where $Y$ is the motive, $\beta$ are the methods, and $\sigma$ is the output section which must
satisfy the propositional coherence conditions $\rho$.
Finally, we can package an inductive algebra over a signature as a telescope
\begin{align*}
&\boxed{\Gamma \vdash S\INDA \tel} \\
&S\INDA = (X : \Delta \to \univ,\ \alpha :: S\A\ (\delta.\ X\ \delta),\ \kappa : \alpha\IND) \,,
\end{align*}
by collecting the carrier $X$, algebra $\alpha$ and induction $\kappa$ all together.

\subsection{Constructing inductive families}\label{sub:constructing-inductive-families}

We now extend intensional \lambdamltt with a type for inductive families, which we
denote $\Mdata_\Delta\ S\ \gamma$. This type defines an inductive family
matching a signature $S$ with indices $\Delta$, together with an inductive
algebra $\gamma$ which `implements' the signature $S$. Notice that this is
different to the usual way that inductive families are defined in type theory,
for example W-types \cite{Abbott2004-va}, where all we need to provide is
a signature.\footnote{
Notice that the data defining a W-type $(A : \univ, B : A \to \univ)$ can be viewed as a kind of signature, where $A$ describes the
operations and their non-recursive parameters, while $B$ describes the arities of the recursive parameters.}
Here, we must also implement the signature and prove
the induction principle by providing $\gamma$, rather than it being `built-in' to
the type theory. For example, if the type theory has W-types, then we can
construct $\gamma$ using an appropriate W-type. In effect, this will later allow us to
translate away inductive definitions to their defined representations.
This leads to the formal definition of a representation:
\begin{definition}
A \emph{representation} of a signature $S$ is an inductive algebra for $S$.
\end{definition}

In \cref{fig:data-rules} we define the $\Mdata$ type, and its corresponding
introduction, elimination, and computation rules.
Constructors form an algebra for the signature $S$ over
$\Mdata_\Delta\ S\ \gamma$, denoted by $\Mctor_S = (\Mctor_{S.O})_{O \in S}$.
Similarly, the eliminator forms a coherent section over the constructor algebra,
which holds definitionally.

\begin{figure}[t]
\begin{mathpar}
\inferrule[\htarget{Data-Form}]
{\Gamma \vdash S \sig \Delta \\ \Gamma \vdash \gamma :: S\INDA \\ \Gamma \vdash \delta :: \Delta}
{\Gamma \vdash \Mdata_\Delta\ S\ \gamma\ \delta \type} \and
\inferrule[\htarget{Data-Intro}]
{O \in S \\ \Gamma \vdash \nu :: O\IN\ (\Mdata_\Delta\ S\ \gamma)}
{\Gamma \vdash \Mctor_{S.O}\ \nu : \Mdata_\Delta\ S\ \gamma\ \nu\OUT} \and
\inferrule[\htarget{Data-Elim}]
{\Gamma,\ \delta :: \Delta,\ \Mdata_\Delta\ S\ \gamma\ \delta \vdash M \type \\
\Gamma \vdash \beta :: \Mctor_S\D\ M \\
\Gamma \vdash \delta :: \Delta \\
\Gamma \vdash x : \Mdata_\Delta\ S\ \gamma\ \delta}
{\Gamma \vdash \Melim_S\ M\ \beta\ \delta\ x : M[\delta, x]} \and
\inferrule[\htarget{Data-Comp}]
{O \in S \\ \Gamma \vdash \nu :: O\IN\ (\Mdata_\Delta\ S\ \gamma) \\
\Gamma,\ \delta :: \Delta,\ \Mdata_\Delta\ S\ \gamma\ \delta \vdash M \type \\
\Gamma \vdash \beta :: \Mctor_S\D\ M}
{\Gamma \vdash \Melim_S\ M\ \beta\ \nu\OUT\ (\Mctor_{S.O}\ \nu) = \beta_O\ (\Melim_S\ M\ \beta \ \$\ \nu) : M[\nu\OUT, \Mctor_{S.O}\ \nu]}
\end{mathpar}
\caption{Rules for data types, constructors and eliminators. We write $O \in S$
to indicate that $O$ is an operation in the signature $S$. We write $\alpha_O$
to extract the telescope element corresponding to operation $O$ from the algebra
$\alpha$ for $S$.}
\label{fig:data-rules}
\end{figure}

One might think, what do we gain by adding $\Mdata$ to the theory? If we can
provide an inductive algebra $\gamma$ for a signature $S$ ourselves, then why
not use $\gamma$ directly? The reason is that by having a primitive for
inductive types, we can take advantage of their properties in an extensional
way. For example, an induction principle suggests that the constructors
corresponding to each method are disjoint. Since constructors $\Mctor$ are
primitive terms in the theory, we can make use of this when formulating a
unification algorithm. Aside from disjointness, we can also rely on other
properties such as injectivity, acyclicity, and `no-confusion'. McBride
\cite{McBride2006-fp} originally explored the properties which arise from the
existence of induction principles, or equivalently, initiality.

For example, if we have an inductive algebra $(\n{N}, \n{zero}_{\n{N}},
\n{succ}_{\n{N}}, \n{elim}_{\n{N}})$ for the natural numbers signature
$\n{NatSig}$ (\cref{ex:nat-sig}), we can prove propositionally that for all $x : \n{N}$,
$\n{zero}_{\n{N}} \neq \n{succ}_{\n{N}}\ x$, by invoking $\n{elim}_{\n{N}}$.
However, the typechecker does not know this fact; it is not derivable as a
definitional equality contradiction. However, it \emph{is} derivable
definitionally that for all $x : \mathbb{N}$, $\Mctor_{\n{zero}} \neq
\Mctor_{\n{succ}}\ x$, where $\mathbb{N} = \Mdata\ {\n{NatSig}}\ (\n{N},
\n{zero}_{\n{N}}, \n{succ}_{\n{N}}, \n{elim}_{\n{N}})$ because the
syntax does not equate $\Mctor_i$ and
$\Mctor_j$ unless $i = j$.

Importantly, the existence of $\Mdata$ enables the use of
\emph{dependent pattern matching} on its inhabitants. Nested pattern matching on
$\mathbb{N}$, for example, can be elaborated to invocations of
$\Melim_{\mathbb{N}}$, which has the expected computation rules as shown in
\cref{fig:data-rules}. Converting dependent pattern matching to eliminators has
been explored in depth by Goguen, McBride and McKinna \cite{Goguen2006-sy}, as
well as by Cockx and Devriese \cite{Cockx2018-bv} in absence of Axiom K.

\subsection{Reasoning about representations} \label{sub:lambdadata}

So far we are able to construct data types using the $\Mdata_\Delta\ S\ \gamma$
type constructor. These data types are themselves implemented in terms of
inductive algebras. However, the rules for data types do not utilise them. We
would like to be able to relate data types to their underlying inductive
algebras. One reason is to avoid unnecessary
computation. If we have a type $X$ that is the carrier of two inductive algebras
$(X, \alpha, \kappa)$ and $(X, \alpha', \kappa')$ for signatures $S$ and
$S'$ respectively, then we can form the data types $D = \Mdata_\Delta\ S\ (X,
\alpha, \kappa)$ and $D' = \Mdata_\Delta\ {S'}\ (X, \alpha', \kappa')$ and make
use of the structural properties of initiality. However, we would also like to
be able to freely convert between $X$, $D$ and $D'$ without incurring any
runtime cost. After all, $D$ and $D'$ are meant to be translated away to their
underlying representation, $X$. This argument can also be made in the context of
theorem proving: sometimes it is easier to prove a property about $D$ or $D'$,
due to their structure, but we should be able to `transport' the property to
$X$.

To make use of these conversions in a computationally-irrelevant manner,
while still retaining the fact that $D$, $D'$ and $X$ are distinct types,
we introduce a modality
$$
\MRepr : \univ \to \univ ,
$$
which takes types to their representations. It comes with two term formers
$\Mrepr$ and $\Munrepr$, which are definitional inverses of each other. We
highlight the main rules of $\MRepr$ in \cref{fig:repr-rules}.

\begin{figure}[H]
\begin{mathpar}
\inferrule[\htarget{Repr-Form}]{\Gamma \vdash A \type}{\Gamma \vdash \MRepr\ A \type} \and
\inferrule[\htarget{Repr-Intro}]{\Gamma \vdash a : A}{\Gamma \vdash \Mrepr\ a : \MRepr\ A} \and
\inferrule[\htarget{Repr-Elim}]{\Gamma \vdash a : \MRepr\ A}{\Gamma \vdash \Munrepr\ a : A} \and
\inferrule[\hypertarget{Repr-Id1}{$\axiom{Repr-Id}_1$}]{\Gamma \vdash a : \MRepr\ A}
{\Gamma \vdash \Mrepr\ (\Munrepr\ a) = a : \MRepr\ A} \and
\inferrule[\hypertarget{Repr-Id2}{$\axiom{Repr-Id}_2$}]{\Gamma \vdash a : A}
{\Gamma \vdash \Munrepr\ (\Mrepr\ a) = a : A} \and
\inferrule[\htarget{Repr-Data}]{\Gamma \vdash S \sig \Delta \\ \Gamma \vdash \gamma :: S\INDA \\ \Gamma \vdash \delta :: \Delta}
{\Gamma \vdash \MRepr\ (\Mdata_\Delta\ S\ \gamma\ \delta) = \gamma.X\ \delta}
\end{mathpar}
\caption{Introduction and elimination forms, as well as computation rules
for the $\MRepr$ modality.}
\label{fig:repr-rules}
\end{figure}

These rules allow us to go between a data type $D = \Mdata_\Delta\ S\ \gamma\
\delta$ and its representation $\gamma.X\ \delta$. In the translation to
extensional \lambdamltt that we are yet to define, this modality is also
translated away. Indeed, its purpose is purely intensional: we do not want to
equate $\Mdata_\Delta\ S\ \gamma$ with $\gamma.X$ because that would render
conversion checking undecidable, but we still want to make use of the fact that
these types are `the same'. In other words, $\MRepr\ A \simeq A$ but not
$\MRepr\ A = A$. All the contextual machinery of general modal type systems
\cite{Gratzer2020-kf} is not necessary here because this modality is fibred over
contexts so it presents as a type former.

One might hope for additional computation rules. For example, the representation
of a constructor should be equal to the underlying algebra element of the
constructor type's representation:\footnote{When we write $\Mrepr\ \nu$ we mean
to apply $\Mrepr$ to all recursive occurences (all places that $\iota$ appears
in the domain of $O$).} \[\Mrepr\ (\Mctor_{S.O}\ \nu) = \gamma.\alpha_O\ (\Mrepr
\ \nu) \,.\] Unfortunately, having this as a computation rule would render
conversion checking undecidable, because if one applies $\mta{unrepr}$ to a term
$\Mrepr\ (\Mctor_{S.O}\ \nu)$ which has already been reduced to its
representation, $\mta{unrepr}\ (\gamma.\alpha_O\ (\Mrepr\ \nu))$, there is no
clear way to decide that this is convertible to $\Mctor_{S.O}\ \nu$ even though
the definitional equality rules would imply that it is (due to
\hyperlink{Repr-Id2}{$\axiom{Repr-Id}_2$}). Nevertheless, we can still postulate this equality
propositionally, and it is justified by the translation step.

We can also ask for compatibility equations between $\MRepr$ and the rest of
the type formers of \lambdamltt; for example $\MRepr\ ((a : A) \times B[a]) = (a
: \MRepr\ A) \times \MRepr\ B[\Munrepr\ a]$, which can hold definitionally
without breaking conversion. These are given for $\univ$, $\Pi$, $\Sigma$, unit
and identity types in the Agda formalisation.


\subsubsection{Subuniverse of concrete types}

As described, the modality $\MRepr$ is not idempotent: $\MRepr\
(\MRepr\ A) = \MRepr\ A$ does not always hold. An inductive algebra used as a
representation of a data type might itself be implemented in terms of another
data type, which again reduces under the action of $\MRepr$. A more principled
approach might be to view the image of $\MRepr$ as a subuniverse of $\univ$. The
restriction
$$
\MRepr : \univ \to \univ_C
$$
targets a universe of `concrete' types $\univ_C < \univ$ closed under all
standard type formers, but without any $\Mdata$ types. We can
then require that all inductive algebras $\gamma$ used in the rule
\hyperlink{Data-Form}{\axiom{Data-Form}} must have a concrete carrier $X : \Delta \to \univ_C$. This
does not limit expressivity because we can always wrap any inductive algebra
carrier with $\MRepr$ to bring it down to $\univ_C$. We do not assume this
additional structure for simplicity, but it might be a useful feature in
practice. For example, it would simplify the transitivity example in
\cref{sub:transitivity}.

\subsection{Translating to extensional \langmltt} \label{sub:translation}

We now define a type- and equality-preserving translation step $\R$ shown in
\cref{fig:translation} from \lambdadata to extensional \langmltt, to be
applied during the compilation process. The extensional flavour of \lambdamltt
involves adding the equality reflection rule
\[
\inferrule[\axiom{Reflect $p$}]{\Gamma \vdash p : a =_A a'}{\Gamma \vdash a \equiv a' : A}
\]
General undecidability of conversion is not a problem because type checking is
decidable for $\lambdadata$\footnote{Not formalised in this paper.} and we
apply this transformation after type checking, on fully-typed terms. The
translation is defined over the syntax of $\lambdadata$ \cite{Boulier2017-cm}
such that definitional equality is preserved, shown in
\cref{fig:translation}. $\R$ replaces data types with their underlying inductive
algebras. We use the notation $\Ty$ and $\Tm$ for types and terms respectively,
with a subscript indicating the language.

\begin{figure}
\begin{minipage}[t]{0.5\textwidth}
\begin{align*}
&\boxed{\R : \Ty_{\langdata}\ \Gamma \to \Ty_{\langmltt}\ \R \Gamma} \\
&\R (\MRepr\ A) = \R A \\
&\R (\Mdata_{\Delta}\ S\ \gamma\ \delta) = \R (\gamma.X)\ \R \delta \\
&\text{(otherwise recurse with $\R$)}
\end{align*}
\end{minipage}%
\begin{minipage}[t]{0.5\textwidth}
\begin{align*}
&\boxed{\R : \Tm_{\langdata}\ \Gamma\ A \to \Tm_{\langmltt}\ \R \Gamma\ \R A} \\
&\R (\Mrepr\ t) = \R t \\
&\R (\Munrepr\ t) = \R t \\
&\R (\Mctor_{S.O}\ \{\gamma\}\ \nu) = \R (\gamma.\alpha_O)\ \R\nu \\
&\R (\Melim_{S}\ \{\gamma\}\ M\ \beta\ \delta\ x)  \\ &\qquad = (\R (\gamma.\kappa)\ \R M\ \R \beta).\sigma\ \R\delta\ \R x \\
&\text{(otherwise recurse with $\R$)}
\end{align*}
\end{minipage}
\caption{Translation of $\lambdadata$ to $\lambdamltt$, replaces data types
with their underlying inductive algebras, and eliminators by the induction
principle provided by representations.}
\label{fig:translation}
\end{figure}

All the mappings above are structurally recursive, demonstrated by the construction
of a  \emph{model} of \lambdadata in the Agda formalisation. The translation is
extended to contexts, substitutions, telescopes and spines pointwise, and
the rest of the syntax is preserved: $\R((a : A) \to B[a]) = (a : \R A) \to (\R
B)[a]$ etc. All $\Mdata$ types are translated to the carriers of their inductive
algebras. Invocations of $\MRepr$ are removed. One can view $\MRepr$ as locally
applying $\R$ to a part of the program. The definitional equality preservation
by $\R$ is shown in \cref{fig:translation-eq}. The isomorphism of
$\Mrepr/\Munrepr$, as well as the rule \hyperlink{Repr-Data}{\axiom{Repr-Data}} are preserved by
metatheoretic reflexivity on the other side, since all representation operators
are erased. Coherence rules for eliminators are preserved by reflecting the
propositional coherence rules provided by their defined representations.

\begin{figure}[H]
\begin{minipage}[t]{0.45\textwidth}
\begin{align*}
&\boxed{\apR^{\Ty\ \Gamma} : A \approx_{\langdata} A' \to \R A \approx_{\langmltt} \R A'} \\
&\R (\axiom{Repr-Data}\ \{S, \gamma, \Delta\}) \\ &\qquad = \axiom{Refl}\ (\R(\gamma.X)\ \R \delta) \\
&\text{(otherwise recurse with $\apR$)}
\end{align*}
\end{minipage}\hspace{\fill}%
\begin{minipage}[t]{0.45\textwidth}
\begin{align*}
&\boxed{\apR^{\Tm\ \Gamma\ A} : a \approx_{\langdata} a' \to \R a \approx_{\langmltt} \R a'} \\
&\R (\axiom{Repr-Id}_1\ \{a\}) = \axiom{Refl}\ \R a \\
&\R (\axiom{Repr-Id}_2\ \{a\}) = \axiom{Refl}\ \R a \\
&\R (\axiom{Data-Comp}\ \{S, O, \gamma, M, \beta, \nu\})  \\
&\quad = \axiom{Reflect}\ (\R (\gamma.\kappa)\ \R M\ \R \beta).\rho_O\ \R\nu \\
&\text{(otherwise recurse with $\R_\approx$)}
\end{align*}
\end{minipage}
\caption{Equality translation of $\lambdadata$ to extensional $\lambdamltt$. This amounts to
an inductive proof that $\R$ preserves equality.}
\label{fig:translation-eq}
\end{figure}

\newcommand{\mech}[1]{\href{#1}{\textsc{agda}}}

\begin{theorem}[\mech{https://github.com/kontheocharis/rep-agda/blob/0c038fa53d05b796570d0a8c0f0e5e06bedef062/TT/Translation.agda\#L19}]
    $\R$ preserves typing and definitional equality.
\end{theorem}

\begin{theorem}[\mech{https://github.com/kontheocharis/rep-agda/blob/e6bd34adaab630f5787c63a95fa86869f6c19da4/TT/Translation.agda\#L124}]
    $\R$ is a left-inverse of the inclusion $i : \lambdamltt
    \hookrightarrow \lambdadata$:
    \[
    \inferrule{\Gamma \vdash a : A}{\Gamma \vdash \R(i(a)) = a : A} \quad \text{(in extensional \lambdamltt)}.
    \]
\end{theorem}

\subsubsection{Inductive types in an empty context}

Basic \lambdamltt with dependent functions, pairs and propositional equality as
we have presented here is not sufficient to construct most inductive algebras in
an empty context. Without W-types or fixpoints, we must postulate the induction
principles we have access to, like the example with GMP integers in
\cref{sec:examples}. In practice, we often want to be able to construct
inductive types `from scratch'. For this, we can extend the base theory
with W-types or something similar. One convenient choice is to extend
the syntax of both the source and target languages of $\R$ with a class of data
types like in \cref{fig:data-rules}, but without requiring them to be
implemented by inductive algebras. The advantage is that now we can opt-in when
we want to represent inductive types specially, but otherwise fall back to some
kind of default implementation. The downside is that now the target still
contains inductive types, though this is okay for compilation purposes if we
choose a `sane default', usually tagged unions containing indirections. This is
the approach we take in \superfluid.

\subsection{Computational irrelevance}\label{sub:irr}

In a compiler, there will be an additional program extraction step from the
target of $\R$ into some simply-typed or untyped language, to be handled by the
code-generation backend. We call this language \lambdaprog, and the transformation by
vertical bars $|x|$. As opposed to $\R$, it might not preserve the definitional
equality of the syntax---we might want to compile two definitionally equal terms
differently. For example, we might not always want to reduce function
application redexes. We will use the \texttt{monospace} font for terms in
\lambdaprog.

\newcommand{\Pid}{\ensuremath{\texttt{\textbackslash x => x}}}

\begin{definition}
    A function $\Gamma \vdash f : (a : A) \to B$, is \emph{computationally irrelevant} if
    $|\R A| = |\R B|$ and $|\R f| = \Pid$.
\end{definition}

\begin{theorem}[\mech{https://github.com/kontheocharis/rep-agda/blob/e6bd34adaab630f5787c63a95fa86869f6c19da4/TT/Lemmas.agda\#L147}]
  \label{thm:repr-inj}
	The type former $\MRepr{}$ is injective up to equivalence, i.e.
	\begin{equation}
	\inferrule{\Gamma \vdash p : \MRepr\ T =_{\univ} \MRepr\ T'}{\Gamma\ \vdash \text{conv}_p : T \simeq T'},
	\end{equation}
	 and if $|-|$ erases internal equality reasoning, $\text{conv}_p$ is computationally irrelevant.
	\begin{proof}
	For the input proof $p$, for $\text{conv}_p$ we have
	$\lambda x.\ \Munrepr_{T'}\ (\mta{coe}\ (\Mrepr\ x)\ p)$
	and for $\text{conv}_p^{-1}$ we have
	$\lambda x.\ \Munrepr_{T}\ (\mta{coe}\ (\Mrepr\ x)\ (\mta{sym}\
	p))$. Both directions map to $\lambda x.\ \mta{coe}\ \_\ x$ by $\R$ which becomes \Pid{} by $|-|$.
	\end{proof}
\end{theorem}

In addition, we can reason about the computational irrelevance of
refinements. Consider extending both source and target languages of $\R$ with
usage-aware `subset' dependent pairs
\[
\inferrule{\Gamma \vdash A \type \\ \Gamma, A \vdash B \type}{\Gamma \vdash \{A \mid B\} \type}
\]
in such a way that $\MRepr$ and $\R$ preserve them, but the extraction step
erases the right component, i.e. $|\{A \mid B\}| = |A|$, $|(x, y)| = |x|$ and
$|\pi_1 x| = |x|$. This can be implemented using quantitative type
theory for example. Suppose we have an inductive family $G = \mta{data}_I\ S_G\
\gamma_G$ over some index type $I$, and an inductive type $F = \mta{data}\ S_F\
\gamma_F$ such that $G$ is represented by a refinement $r : F \to I$, meaning
\[
    \gamma_G = (\lambda i.\ \{ f : F \mid r\ f \equiv_I i \}, \alpha, \kappa) \,.
\]
Then, we can construct computationally irrelevant functions
\begin{align*}
&\begin{aligned}
&\mta{forget}_i : G\ i \to F \\
&\mta{forget}_i = \lambda g.\ \pi_1\ (\mta{repr}\ g)
\end{aligned} \quad \begin{aligned}
&\mta{remember} : (f : F) \to G\ (r\ f) \\
&\mta{remember} = \lambda x.\ \mta{unrepr}\ (x, \mta{refl}) \,.
\end{aligned}
\end{align*}
By reasoning similar to \cref{thm:repr-inj}, $|\R\ \mta{forget}_i| = |\R\
\mta{remember}| = \Pid$.

\section{Implementation}\label{sec:implementation}

\superfluid is a programming language with dependent types with quantities,
inductive families and data representations. Its compiler is written in Haskell
and the compilation target is JavaScript. After prior to code generation, the
$\R$ transformation is applied to the elaborated core program, which erases all
inductive constructs with defined representations. Then, a JavaScript program is
extracted, erasing all irrelevant data by usage analysis similarly to Idris 2.
As a result, with appropriate postulates in the prelude, we are able to
represent $\datalab{Nat}$ as JavaScript's \texttt{BigInt}, and $\datalab{List}\
T$/$\datalab{SnocList}\ T$/$\datalab{Vec}\ T\ n$ as JavaScript's arrays with the
appropriate index refinement, such that we can convert between them without any
runtime overhead. The syntax of \superfluid very closely mirrors the syntax
given in the first half of this paper. It supports global definitions, inductive
families, as well as postulates. Users are able to define custom representations
for data types using $\defkwd{repr} - \defkwd{as}\ -$ blocks.

Currently we do not require proofs of eliminator coherence, but they are
straightforward to add. We also treat the rule for representing constructors
($\Mrepr\ (\Mctor_{S.O}\ \nu) = \gamma.\alpha_O\ (\Mrepr\ \nu)$) as definitional
in the implementation, at the cost of breaking decidability of equality, but
with the benefit of fewer manual transports. \superfluid also supports the definition of
representations for global functions in addition to inductive families. This is a
simple symbol-replacement mechanism (like Idris's \texttt{\%transform} pragma)
so that we can still take advantage of inductively defined functions---such as addition
on natural numbers---for theorem proving, but use the optimised primitive addition
when generating code.

We are currently working on adding
dependent pattern matching that is elaborated to internal eliminators, so that
we can take advantage of the structural unification rules for data types
\cite{McBride2006-fp}. We have written some of the examples in this paper in
\superfluid, which can be found in the \texttt{examples} directory. Overall the
implementation is a proof of concept, but we expect that our framework can be
implemented in an existing language.

\section{Related work}

Using inductive types as a form of abstraction was first explored by Wadler
\cite{Wadler1987-zp} through \emph{views}. The extension to dependent types
was developed by McBride and McKinna \cite{Mcbride2004-fd}, as part of the
Epigram project. Our system differs from views in the computational content
of the abstraction; even with deforestation \cite{Wadler1990-yo} views are not
always zero-cost, but representations are.
Atkey \cite{Atkey2011-ex} shows how to generically derive inductive types which
are refinements of other inductive types. This work could be integrated in our
system to automatically generate representations for refined data types.
Zero-cost data reuse when it comes to refinements of inductive types has been
explored in the context of Church encoding in Cedille \cite{Diehl2018-ba}, but
does not extend to custom representations.

Work by Allais \cite{Allais2023-pf,Allais2023-zq} uses a combination of views,
erasure by quantitative type theory, and universes of flattened data types to
achieve performance improvements when working with serialised data in Idris 2.
Our approach differs because we have access to `native' data representations, so
we do not need to rely on encodings. Additionally, they rely on partial evaluation
to erase their views, which does not always fire. On the topic of memory layout
optimisation, Baudon \cite{Baudon2023-cy} develops Ribbit, a DSL for the
specification of the memory representation of algebraic data types, which can
specify techniques like struct packing and bit-stealing. To our knowledge
however, this does not provide control over the indirection introduced by
\emph{inductive} types.

Dependently typed languages with extraction features, including Idris 2 \cite{idris-extraction}, Rocq
\cite{coq-extraction} and Agda \cite{agda-extraction}, have some overlapping
capabilities with our approach, but they do not provide any of the correctness
guarantees. Optimisation tricks such as the Nat-hack, and its generalisation to
other types, can emulate a part of our system but are unverified and special
casing in the compiler. Since the extended abstract version of this paper, an
optimisation was merged into Idris 2 \cite{idris-pr} to erase the forgetful and
recomputation functions for reindexing list/maybe/number-like types.
There is also demand for this kind of optimisation in Agda
\cite{agda-issue}.


\section{Future work}


There are elements of our formalisation which should be developed further. We
did not formulate normalisation and decidability of equality for $\lambdadata$,
which is needed for typechecking. We have implemented a
normalisation-by-evaluation \cite{Abel2013-fq} algorithm used in \superfluid, but have only
sketched that it has the desired properties. On the practical side, we have not
explored examples of representations in great detail. Once the implementation of
$\superfluid$ is more complete, we would like to explore more sophisticated and
complete examples, along with compelling benchmarks. We would also like to
describe \superfluid's feature of representing global function definitions more
formally in the future.

As a next step we aim to expand the class of theories we consider, in particular
to include quotient-induction. Representations for quotient-inductive types in
could give rise to ergonomic ways of computing with more `traditional' data
structures such as hash maps or binary search trees. We could program
inductively over these structures but extract programs without redundancy in
data representation. Additionally, quotient-inductive types could be a good
candidate for improving typechecking ergonomics by deciding their equational
theories, similar to Frex \cite{Allais2023-rg}. Such a system could
apply certain representations during typechecking rather than code generation,
to solve equations involving free variables through a normalisation-by-evaluation
procedure.


We would like to further explore the relationship of the $\mta{Repr}$ modality with
general systems for defining modalities, such as \emph{multi-modal type theory}
by Gratzer \cite{Gratzer2020-kf}. Additionally, we expect that metaprogramming with
representations is most naturally done in the context of \emph{two-level type theory} (2LTT)
\cite{Kovacs2022-vb}. We would like to explore the embedding of $\lambdadata$
in a two-level type theory, where signatures become types in the meta-fragment.
Then we could develop various methods of generating representations internally,
for example through algebraic ornaments, without needing to laboriously prove
induction principles by hand. The translation step in \cref{sub:translation}
can already be viewed as a kind of staging procedure, and could be
integrated with the one of 2LTT.

\section{Conclusion}

This paper addresses some of the inefficiencies of inductive families in
dependently typed languages by introducing custom runtime representations that
preserve logical guarantees and simplicity of the surface language while
optimising performance and usability. These representations are formalised as inductive
algebras, and come with a framework for reasoning about them: provably zero-cost
conversions between original and represented data.
The compilation process guarantees erasure of abstraction layers, translating
high-level constructs to their defined implementations. Our hope is that by
decoupling logical structure from runtime representation, type-driven
correctness can be leveraged further without great sacrifices in performance.

\subsubsection{Acknowledgements}
We thank the anonymous reviewers for their feedback that
improved the quality of this paper, Guillaume Allais for helpful
comments on the extended abstract, as well as Nathan Corbyn, Ellis
Kesterton, and Matthew Pickering for interesting discussions surrounding it.

\def\UrlBreaks{\do\/\do-}

\bibliographystyle{splncs04}
\bibliography{bib}

\clearpage
\section{Appendix}

\allowdisplaybreaks

\paragraph{Implementation}

The implementation of \superfluid can be found at \url{https://github.com/kontheocharis/superfluid}.

\paragraph{Formalisation} \label{app:formalisation}

The Agda formalisation of the developments of this paper at \url{https://github.com/kontheocharis/rep-agda}.

\subsection{Definition of \lambdamltt}

For reference, we define Martin-Löf type theory with $\univ :
	\univ$, $\Pi$, $\Sigma$, propositional equality and unit. We omit the
definition of the substitution calculus and equality coercions
(see \cite[5.1.2]{Castellan2019-qo}). We use de-Brujin indices, keeping weakening
implicit, and abuse notation for substitutions of terms: $A[t]$ for $A[\mta{id},
			t]$.

\newcommand{\refl}{\mta{refl}}
\newcommand{\fst}{\mta{fst}}
\newcommand{\snd}{\mta{snd}}
\newcommand{\pair}{\mta{pair}}
\newcommand{\app}{\mta{app}}
\newcommand{\lam}{\mta{lam}}
\newcommand{\J}{\mta{J}}

\begin{figure}[h]
	\begin{mathpar}
		\inferrule[Univ-Form]
		{ }
		{\Gamma \vdash \univ \type}

		\inferrule[El-Form]
		{\Gamma \vdash a : \univ}
		{\Gamma \vdash \El a \type}

		\inferrule[Univ-Intro]
		{\Gamma \vdash A \type}
		{\Gamma \vdash \code A : \univ}

		\inferrule[Pi-Form]
		{\Gamma \vdash A \type \\ \Gamma, x : A \vdash B \type}
		{\Gamma \vdash (x : A) \to B \type}

		\inferrule[Pi-Intro]
		{\Gamma, x : A \vdash b : B}
		{\Gamma \vdash \lambda\ x.\ b : (x : A) \to B}

		\inferrule[Pi-Elim]
		{\Gamma \vdash f : (x : A) \to B \\ \Gamma \vdash a : A}
		{\Gamma \vdash f\ a : B[a]}

		\inferrule[Eq-Form]
		{\Gamma \vdash A \type \\ \Gamma \vdash a : A \\ \Gamma \vdash b : A}
		{\Gamma \vdash a \equiv_A b \type}

		\inferrule[Eq-Intro]
		{\Gamma \vdash a : A}
		{\Gamma \vdash \refl\ a : a \equiv_A a}

		\inferrule[Eq-Elim]
		{\Gamma \vdash A \type \\
			\Gamma, a : A, b : A, a \equiv_A b \vdash P \type \\
			\Gamma, a : A \vdash r : P[a, a, \refl\ a] \\
			\Gamma \vdash a : A \\ \Gamma \vdash b : A \\ \Gamma \vdash p : a \equiv_A b}
		{\Gamma \vdash \J\ P\ d\ p : P[a, b, p]}

		\inferrule[Unit-Form]
		{ }
		{\Gamma \vdash \top \type}

		\inferrule[Unit-Intro]
		{ }
		{\Gamma \vdash \mta{tt} : \top}

		\inferrule[Sigma-Form]
		{\Gamma \vdash A \type \\ \Gamma, x : A \vdash B \type}
		{\Gamma \vdash (x : A) \times B \type}

		\inferrule[Sigma-Intro]
		{\Gamma \vdash a : A \\ \Gamma \vdash b : B[a]}
		{\Gamma \vdash (a, b) : (x : A) \times B}

		\inferrule[Sigma-Elim-Fst]
		{\Gamma \vdash p : (x : A) \times B}
		{\Gamma \vdash \fst\ p : A}

		\inferrule[Sigma-Elim-Snd]
		{\Gamma \vdash p : (x : A) \times B}
		{\Gamma \vdash \snd\ p : B[\fst\ p]}
	\end{mathpar}
	\caption{Typing rules for \lambdamltt.}
\end{figure}

\begin{figure}[h]
	\begin{minipage}[t]{0.5\textwidth}
		\begin{align*}
			 & \boxed{\text{Universes}}      \\
			 & {\El (\code A) = A}           \\
			 & {\code (\El t) = t}           \\[1em]
			 & \boxed{\text{$\Pi$ types}}    \\
			 & { (\lambda\ x.\ b)\ a = b[a]} \\
			 & { \lambda\ x.\ (f\ x) = f }
		\end{align*}
	\end{minipage} \qquad
	\begin{minipage}[t]{0.5\textwidth}
		\begin{align*}
			 & \boxed{\text{Equality and unit}} \\
			 & { \J\ P\ r\ (\refl\ a) = r\ a }  \\
			 & { x = \mta{tt}}                  \\[1em]
			 & \boxed{\text{$\Sigma$ types}}    \\
			 & { \fst\ (a, b) = a}              \\
			 & { \snd\ (a, b) = b}              \\
			 & { (\fst\ p, \snd\ p) = p }
		\end{align*}
	\end{minipage}
	\caption{Definitional equality rules for \lambdamltt, omitting substitution rules
		such as $(\El a) [\sigma] = \El (a[\sigma])$.}
\end{figure}

\subsection{Definition of \lambdadata}

The language \lambdadata is the extension of \lambdamltt. by the rules in
\cref{fig:data-rules,fig:repr-rules}. Below we present some additional
definitional equality rules of $\MRepr$ that make it commute with most of the
syntax, as well as some propositional equalities that are justified by the translation $\R$.

\begin{figure}[H]
	\begin{minipage}[t]{0.4\textwidth}%
		\begin{align*}
			 & \boxed{\text{$\eta$ rules}}                               \\
			 & \Munrepr\ (\Mrepr\ t) = t                                 \\
			 & \Mrepr\ (\Munrepr\ t) = t                                 \\[1em]
			 & \boxed{\text{Stability under substitution}}               \\
			 & \Mrepr\ {(t[\sigma])} = (\Mrepr\ {t})[\sigma]             \\
			 & \Munrepr\ {(t[\sigma])} = (\Munrepr\ {t})[\sigma]         \\
			 & \MRepr\ {(T[\sigma])} = (\MRepr\ {T})[\sigma]             \\[1em]
			 & \boxed{\text{Compatibility with $\Pi$ types}}             \\
			 & \MRepr\ {((x : A) \to B)} = (x : A) \to \MRepr\ B         \\
			 & \Mrepr\ {(\lambda\ x.\ b)} = \lambda\ x.\ (\Mrepr\ b)     \\
			 & \Munrepr\ {(\lambda\ x.\ b)} = \lambda\ x.\ (\Munrepr\ b) \\
			 & \Mrepr\ (f\ a) = (\Mrepr\ f)\ a                           \\
			 & \Munrepr\ (f\ a) = (\Munrepr\ f)\ a
		\end{align*}
	\end{minipage}\qquad
	\begin{minipage}[t]{0.5\textwidth}%
		\begin{align*}
			 & \boxed{\text{Compatibility with universes}}                                   \\
			 & \MRepr\ {\univ} = \univ                                                       \\
			 & \Mrepr\ {(\Code A)} = \Code A                                                 \\
			 & \Munrepr\ {(\Code A)} = \Code A                                               \\[1em]
			 & \boxed{\text{Compatibility with equality}}                                    \\
			 & \MRepr\ (a \equiv_A b) = \Mrepr\ a \equiv_{\MRepr A} \Mrepr\ b                \\
			 & \Mrepr\ (\mta{refl}\ a) = \mta{refl}\ (\Mrepr\ a)                             \\
			 & \Munrepr\ (\mta{refl}\ a)  = \mta{refl}\ (\Munrepr\ a)                        \\
			 & \Mrepr\ (\mta{J}\ P\ d\ p) = \mta{J}\ (\mta{Repr}\ P)\ (\mta{repr}\ d)\ p     \\
			 & \Munrepr\ (\mta{J}\ (\mta{Repr}\ P)\ d\ p) = \mta{J}\ P\ (\mta{unrepr}\ d)\ p \\[1em]
			 & \boxed{\text{Compatibility with eliminators}}                                 \\
			 & \Mrepr\ (\mta{elim}_S\ M\ \beta\ \delta\ x)                                   \\ &\qquad = \mta{elim}_S\ (\MRepr\ M)\ (\Mrepr\ \beta)\ \delta\ x \\
			 & \Munrepr\ (\mta{elim}_S\ (\MRepr\ M)\ \beta\ \delta\ x)                       \\ &\qquad = \mta{elim}_S\ M\ (\Munrepr\ \beta)\ \delta\ x
		\end{align*}
	\end{minipage}%
	\caption{(\mech{https://github.com/kontheocharis/rep-agda/blob/e6bd34adaab630f5787c63a95fa86869f6c19da4/TT/Repr.agda\#L41})\
		Definitional compatibility rules for $\MRepr$. Similar rules are given for
		$\Sigma$ and $\top$ in the formalisation. In this version, $\MRepr$ only applies to codomains
		of functions which aligns with the substitution rule. However, it is also
		possible to formulate it as \\ $\MRepr\ ((x : A) \to B) = (x : \MRepr\ A) \to \MRepr\ B[\Munrepr\ x]$.}
	\label{fig:lambdaind-repr-coherence-pi-univ}
\end{figure}

\begin{figure}[H]
	\vspace{2em}
	\begin{align*}
		 & \mta{repr-ctor}_{S.O}\ \{\gamma\}\ \nu :                \\ & \qquad \Mrepr\ (\Mctor_{S.O}\ \{\gamma\} \nu) \equiv \gamma.\alpha_O\ (\Mrepr\ \nu) \\[1em]
		 & \mta{elim-equiv}_{S}\ \{\gamma\}\ M\ \beta\ \delta\ x : \\   & \qquad
		    \mta{elim}_{S}\ \{\gamma\}\ M\ \beta\ \delta\ x \equiv \gamma.\kappa\ (\delta.\ x.\ M[\delta, \Munrepr\ x])\ (\Mrepr^*\ \beta)\ \delta\ (\Mrepr\ x)
	\end{align*}
	\caption{Additional propositional equalities for $\MRepr$ on constructors and eliminators. Here, $\Mrepr^*$ applies $\MRepr$ on all the recursive
		arguments of a displayed algebra. The rule $\mta{elim-equiv}_{S}\ \{\gamma\}\ M\ \beta\ \delta\ x$ is also derivable internally by case analysis
		on $x$.}
\end{figure}

\end{document}